\documentclass[sigconf]{acmart}
\usepackage{multirow} % we use Times as the main font
\usepackage{subfig}
\usepackage{graphicx}
\usepackage{wrapfig}
\usepackage{enumitem}
\usepackage[ruled]{algorithm2e}
\usepackage{algorithmic}
\usepackage{amsmath}
\usepackage{bm}
\usepackage{tabu}                      % only used for the table example
\usepackage{booktabs}                  % only used for the table example
\usepackage{lipsum}                    % used to generate placeholder text
\usepackage{mwe}                       % used to generate placeholder figures
\usepackage{amsmath}

\usepackage{xcolor} % 
\newif\ifshowrevs
\showrevstrue  
\ifshowrevs
  \newcommand{\rev}[1]{\textcolor{black}{#1}}
\else
  \newcommand{\rev}[1]{#1}
\fi

\AtBeginDocument{%
  \providecommand\BibTeX{{%
    \normalfont B\kern-0.5em{\scshape i\kern-0.25em b}\kern-0.8em\TeX}}}

\begin{document}

\title[Bend It, Aim It, Tap It: Designing an On-Body Disambiguation Mechanism for Curve Selection in Mixed Reality]{Bend It, Aim It, Tap It: Designing an On-Body Disambiguation Mechanism for Curve Selection in Mixed Reality}


\author{Xiang Li}
\orcid{0000-0001-5529-071X}
\email{xl529@cam.ac.uk}
\affiliation{%
  \institution{University of Cambridge}
  \city{Cambridge}
  \country{United Kingdom}
}

\author{Per Ola Kristensson}
\orcid{0000-0002-7139-871X}
\email{pok21@cam.ac.uk}
\affiliation{%
  \institution{University of Cambridge}
  \city{Cambridge}
  \country{United Kingdom}
}

\renewcommand{\shortauthors}{Xiang Li and Per Ola Kristensson}


\begin{abstract}
Object selection in Mixed Reality (MR) becomes \rev{particularly} challenging in dense or occluded environments, where traditional mid-air ray-casting often leads to ambiguity and reduced precision. We present two complementary techniques: \rev{(1)} a real-time Bézier Curve selection paradigm guided by finger curvature, \rev{enabling expressive one-handed trajectories}, and \rev{(2)} an on-body disambiguation mechanism that projects the four nearest \rev{candidates} onto the user’s forearm via proximity-based mapping. Together, these techniques combine flexible, user-controlled selection with tactile, proprioceptive disambiguation. We evaluated their independent and joint effects in a $2 \times 2$ within-subjects study ($N = 24$), crossing interaction paradigm (Bézier Curve vs. Linear Ray) with interaction medium (Mid-air vs. On-body). Results show that on-body disambiguation significantly reduced selection errors and physical demand while improving perceived performance, \rev{hedonic quality}, and user preference. Bézier input \rev{provided effective access to occluded targets but incurred longer task times and greater effort under some conditions}. We conclude with \rev{design implications for integrating curved input and on-body previews} to support precise, adaptive selection in immersive environments.
\end{abstract}



% \received{\today}

\begin{teaserfigure}
  \centering
  \includegraphics[width=0.85\linewidth]{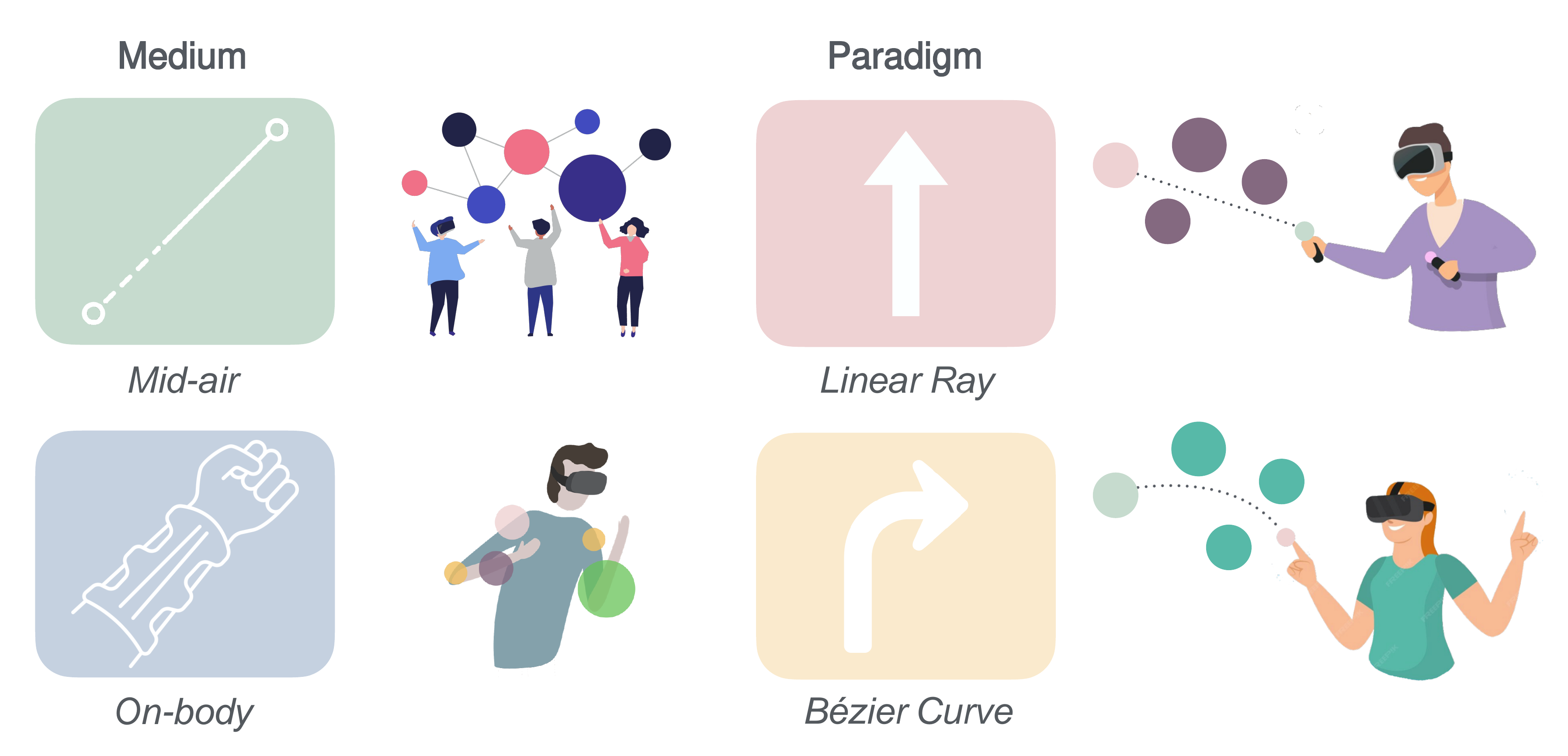}
  \caption{Overview of our $2 \times 2$ study design, crossing two interaction media (Mid-air and On-body) with two interaction paradigms (Linear Ray and Bézier Curve)~\cite{li2025optimizing}. We investigate how the interaction medium and interaction paradigm jointly affect selection performance, effort, and user experience in virtual environments.}
  \Description{A 2×2 study design with two interaction media—Mid-Air and On-Body—and two interaction paradigms—Linear Ray and Bézier Curve. Illustrations show a VR user performing selection either in mid-air or on their forearm, using either a straight ray or a curved Bézier trajectory.}
  \label{fig:teaser}
\end{teaserfigure}

\maketitle

\section{Introduction}
Object selection is a fundamental interaction task in Mixed Reality (MR), yet it becomes particularly challenging in visually dense or spatially occluded environments. Mid-air pointing techniques, such as Linear Ray-casting~\cite{bowman_evaluation_1997, pierce_voodoo_1999}, are widely adopted due to their intuitiveness and compatibility with distant targets. However, they often suffer from selection ambiguity, limited precision, and a lack of tactile feedback, specially when multiple targets are closely spaced or partially occluded~\cite{jr_3d_2017}. As MR applications scale to more complex scenes, there is a growing need for selection techniques that offer both flexibility and disambiguation support.

To address these limitations, prior work has explored two promising directions. The first involves enhancing the expressiveness of the selection path. Curved trajectories, such as Bézier curves, offer users the ability to reach around occluding objects or select targets in cluttered spaces without physically repositioning~\cite{olwal_flexible_2003, steinicke_object_2004, lu_investigating_2020}. For example, the Flexible Pointer~\cite{olwal_flexible_2003} enabled Bézier Curve-based selection by combining two-hand input to modulate curve shape, though it required complex hardware coordination. The Improved Virtual Pointer~\cite{steinicke_object_2004} took a system-driven approach, bending rays toward nearby targets using proximity, but reduced user control over the trajectory. While these techniques demonstrated the value of curved input, they either lacked user-driven control or imposed cumbersome setups. \rev{Our preliminary work~\cite{li2025optimizing} also explored Bézier-based selection with on-body projection but only proposed the initial design with . This paper substantially extends that work with a full implementation and a comprehensive factorial user study.}

The second direction leverages the user’s body as an interaction surface~\cite{harrison_skinput_2010, gustafson_imaginary_2011, yu_blending_2022}. On-body input offers passive haptic feedback and proprioceptive grounding, which can improve selection clarity and reduce cognitive load~\cite{simeone2014feet}. Yet, on-body disambiguation has rarely been integrated with curved input, and it remains unclear whether a single on-body strategy can support both expressive and conventional pointing paradigms.

In this paper, we bridge two strands of MR selection research (i.e., trajectory expressiveness and target disambiguation) by introducing an integrated system that combines real-time curved input with tactile, proprioceptive support. We propose two complementary techniques: (1) a Bézier Curve selection paradigm driven by index finger curvature, enabling expressive, user-controlled trajectories with one hand; and (2) a proximity-based on-body projection mechanism that maps the nearest selectable targets onto the user’s forearm, allowing direct disambiguation via the non-dominant hand. \rev{The number of projected candidates (i.e., 4) is motivated by physical constraints of the forearm and prior findings on on-body menus~\cite{li2024onbodymenu,azai_tap-tap_2018,tsimbalistaia2025body}}

To evaluate their effectiveness and interplay, we conducted a controlled user study ($N = 24$) using a $2 \times 2$ within-subjects design. We varied \textsc{Interaction Paradigm} (Linear Ray vs. Bézier Curve) and \textsc{Interaction Medium} (Mid-air vs. On-body), resulting in four interaction conditions. This factorial design allowed us to examine the independent and combined effects of curve-based input and on-body disambiguation.

Our results show that on-body disambiguation significantly reduces selection errors, lowers physical demand, and improves perceived performance and user preference across paradigms. The Bézier paradigm enables flexible access to occluded targets but introduces trade-offs in task time and physical effort, especially in on-body conditions. Together, these findings reveal the synergies and tensions between on-body disambiguation and trajectory expressiveness in MR selection. Our contributions are as follows:
\begin{itemize}
  \item We introduce a real-time Bézier Curve generation technique driven by finger curvature, enabling expressive and controllable curved selection in immersive environments.
  \item We propose a proximity-based on-body projection mechanism that supports tactile disambiguation by mapping up to four nearby targets onto the user’s forearm.
  \item We empirically demonstrate that on-body disambiguation improves accuracy, reduces workload, and enhances user preference, while Bézier input offers spatial flexibility with measurable trade-offs.
\end{itemize}

\section{Related Work}
\subsection{Body-Centric Interaction}

On-body interaction techniques use the human body as an interactive platform, significantly enhancing immersion and proprioception in virtual reality (VR) environments~\cite{harrison_-body_2012,bergstrom_human--computer_2019,coyle_i_2012,floyd_mueller_limited_2021,mueller_towards_2023}. These approaches improve accuracy in eyes-free targeting, owing to the natural familiarity of users with their own body dimensions~\cite{weigel_skinmarks_2017,gustafson_imaginary_2010}. Extensive research has been conducted on utilizing various body parts (e.g., arms, palms, skin) as interfaces for VR interactions~\cite{chatain_digiglo_2020,dezfuli_palmrc_2014,mistry_wuw_2009,wang2025handows}. For example, Skinput~\cite{harrison_skinput_2010} and Touché~\cite{sato_touche_2012} demonstrate advanced gesture recognition capabilities by detecting acoustic and capacitive signals directly from the skin.

Further studies have assessed the effectiveness of interfaces anchored to the non-dominant arm for precise pointing tasks in VR, as explored by Li et al.~\cite{li_armstrong_2021}. The development of body-centric selection and manipulation techniques such as the Hand Range Interface~\cite{xu_hand_2018}, Swarm Manipulation~\cite{li2023swarm,li2024swarm}, BodyLoci~\cite{fruchard_impact_2018}, and BodyOn~\cite{yu_blending_2022} has opened new possibilities for enhancing mid-air interactions using the human body itself as an interface. Additionally, on-body menus such as the Tap-tap Menu~\cite{azai_tap-tap_2018} and PalmGesture~\cite{wang_palmgesture_2015} have been pivotal in exploring how visual and tactile cues can be effectively utilized to navigate these innovative user interfaces~\cite{li2024onbodymenu}. These investigations highlight the significant potential of on-body interaction to create more engaging VR experiences.

\subsection{Mid-Air Interaction}

Mid-air interaction is a prominent feature in modern headset-based VR systems, allowing users to interact with digital content in virtual environments through gestures and movements. These interactions are often mediated by game controllers or directly through hand movements~\cite{cornelio_martinez_agency_2017,koutsabasis_empirical_2019,song_hotgestures_2023}. Renowned for its straightforward approach, mid-air interaction is particularly adept in 3D spaces due to its versatile input capabilities~\cite{jr_3d_2017}. Users need first to hover the hand over the virtual object and then perform a grab gesture to select the object. However, it is also criticized for its lack of precision~\cite{arora_experimental_2017,mendes2019survey,argelaguet_survey_2013}, the potential for user fatigue~\cite{xu2020exploring,hinckley_survey_1994}, and the absence of tactile feedback~\cite{fang_retargeted_2021}, which can diminish the overall user experience. Jannat et al.~\cite{jannat2024exploring} investigated transitioning AR user interfaces from mid-air to on-body and physical surfaces, finding that users preferred semi-automatic transitions for balancing control and automation.

In response to these challenges, researchers have explored methods to improve the usability and broaden the interaction vocabulary of mid-air systems. Studies have considered more relaxed input methods, such as adopting an arms-down posture to reduce fatigue~\cite{brasier2020arpads,liu2015gunslinger}. In addition, computational techniques have been employed to improve input precision, for example, by developing models that optimize the selection distribution~\cite{yu_modeling_2019,yu_modeling_2023}. These advancements aim to mitigate the limitations of mid-air interactions while leveraging their inherent benefits in immersive environments. 

\subsection{Selection Techniques for VR}

\begin{figure*}[ht]
    \centering
    \includegraphics[width=\linewidth]{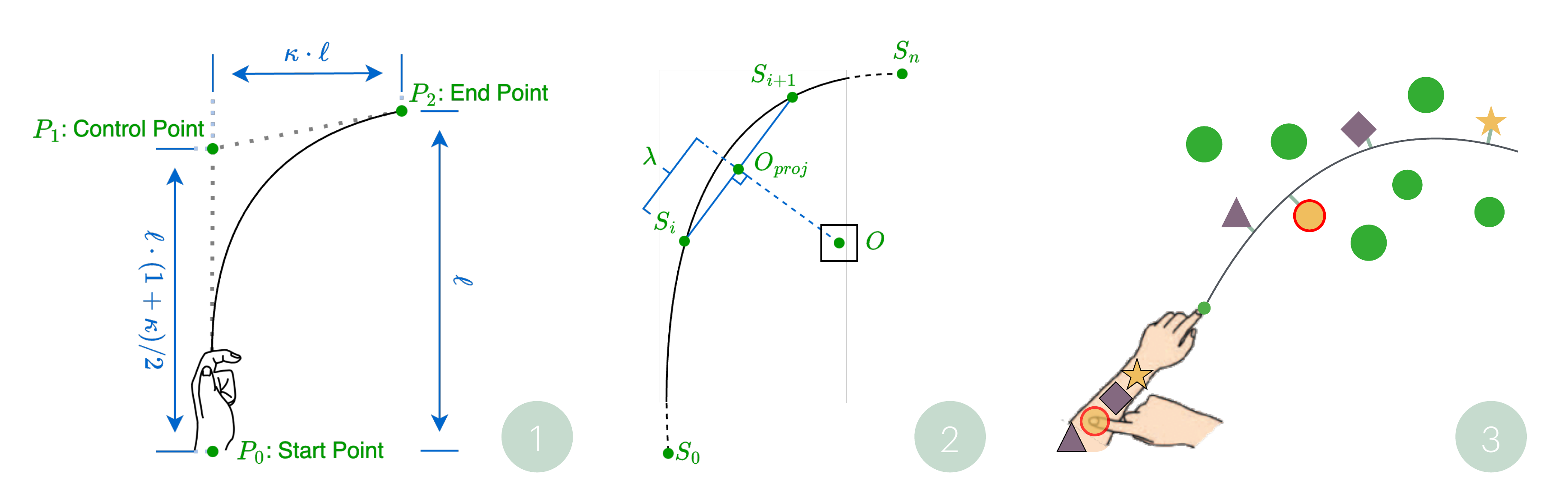}
    \caption{(1) A Bézier Curve defined by start point \( P_0 \), control point \( P_1 \), and end point \( P_2 \), using parameters such as curvature \( \kappa \) and length \( \ell \). (2) Example of how the proximity-matching mechanism works, demonstrating the calculation of the shortest distance from an object \( O \) to points on the curve, specifically between \( S_i \) and \( S_{i+1} \), and the projection \( O_{proj} \). (3) A schematic representation of using the on-body Bézier Curve with the proximity-matching mechanism, where the closest four objects are selected and projected onto the user's forearm.}
    \Description{Three-part figure. (1) A Bézier curve is defined by a start point, control point, and end point, with annotations showing curvature and length. (2) Diagram illustrating the proximity-matching mechanism, showing an object and the shortest distance to a segment of the curve between two sampled points. (3) Illustration of the on-body Bézier curve interaction, where a user selects virtual objects based on proximity and the closest four objects are projected onto their forearm.}
    \label{fig:merged_system}
\end{figure*}

Recently, Yu et al.~\cite{yu2024object} presented a systematic literature review on object selection techniques for VR. Ray-casting is one of the most widely adopted techniques for object selection in virtual reality (VR), typically involving the projection of a ray from the hand, head, or gaze to select intersected objects~\cite{hinckley_survey_1994, pierce_image_1997}. While intuitive and scalable for distant targets, traditional ray-casting often suffers from ambiguity and precision issues, especially in cluttered environments or when selecting small, nearby, or occluded objects~\cite{grossman_design_2006, yu_fully-occluded_2020}. Hand tremors, unstable rays, and overlapping targets can further reduce selection accuracy~\cite{olsen_laser_2001}.

To address these challenges, several strategies have been proposed. Two-step selection methods first narrow down candidates and then use a secondary refinement mechanism~\cite{azai_tap-tap_2018, lediaeva_evaluation_2020}. Target snapping, cone-based assistance~\cite{steed_3d_nodate}, and depth-based control rays~\cite{grossman_design_2006} aim to disambiguate overlapping targets. Bubble mechanisms improve the usability and efficiency of ray-casting techniques~\cite{cashion_dense_2012,grossman_bubble_2005}, while the RayCursor introduced a movable cursor along the ray axis that selects the closest target based on proximity, combining the benefits of ray-casting and bubble selection~\cite{baloup2019raycursor}. \rev{Beyond general enhancements, several techniques focus on dense or occluded environments. Blanch and Ortega~\cite{blanch2011benchmarking} showed that distractor density strongly impacts performance. Starfish~\cite{wonner2012starfish} used a starfish-shaped cursor to guide selection of clustered targets. Waugh et al.~\cite{waugh2025everything} combined area cursors with adaptive control–display gain to improve speed and accuracy in dense touchless input.}

Alternative enhancements include predictive models and system-driven ray bending, such as the Flexible Pointer~\cite{olwal_flexible_2003} and Improved Virtual Pointer~\cite{steinicke_object_2004}, \rev{which employed Bézier curves to bend rays toward occluded targets}, and SenseShapes~\cite{olwal_senseshapes_2003}, which automatically adjust trajectories to improve selection. However, many such techniques reduce user agency by introducing system-controlled behavior. Our approach builds on this space by enabling user-driven, real-time curvature through finger flexion, while offering an on-body disambiguation layer via on-body input.

Despite the breadth of ray-based selection research~\cite{kopper_rapid_2011, looser_evaluation_2007, vanacken_exploring_2007}, challenges remain in balancing flexibility, disambiguation, and user control. \rev{Compared to prior dense-selection and Bézier-based approaches, our work uniquely combines user-controlled curved trajectories with proprioceptive, on-body feedback, addressing both precision and disambiguation in cluttered VR environments.}

\section{Technical Design}

In this section, we outline the technical design of our solutions, specifically focusing on (1) the generation of Bézier Curves from user gestures and (2) a proximity-matching mechanism optimized for real-time calculation. We explore the transformation of user gesture data into Bézier Curve parameters and the real-time projection of objects near the curve onto a predetermined on-body surface (i.e., the forearm).

\subsection{Bézier Curve Formulation}

We chose the Bézier Curve for curve fitting because it has been widely used in computer graphics and has proven ergonomic benefits that simplify interaction~\cite{olwal_flexible_2003,lu_investigating_2020,de_haan_intenselect_nodate,steinicke_object_2004,li2025optimizing}. A quadratic Bézier Curve, integral to our design, is defined by two endpoints (i.e., start point and end point) and a single control point: $\bm{P}_{\text{0}}$, $\bm{P}_{\text{1}}$, and $\bm{P}_{\text{2}}$~\cite{bohm1984survey, Post1989CurvesAS}. The mathematical expression of the curve is:

\begin{equation}
  \bm{B}(t) = \left(1 - t\right)^2 \bm{P}_0 + 2\left(1 - t\right)t \bm{P}_1 + t^2 \bm{P}_2, \quad \left(0 \leq t \leq 1\right)
\end{equation}

where \( t \) is a parameter that determines the position along the curve, with \( t = 0 \) at the start point and \( t = 1 \) at the end point.

\subsection{Gesture to Bézier Mapping}

In our system, the curvature of a virtual ray is controlled by the flexion of the user’s left index finger. This flexion dynamically adjusts the Bézier Curve parameters, allowing the system to modulate the ray’s curvature in real time. Specifically, the length between the fingertip and the wrist is measured to determine the degree of finger flexion. This distance modulates the curvature parameter $\kappa$, which directly influences the shape of the Bézier Curve. While finger flexion primarily drives the curvature, users may also naturally rotate their wrist during interaction. This additional degree of freedom allows the curve to bend in multiple directions, mitigating the limitations of single-axis finger motion and expanding the expressive range of the trajectory.

\begin{figure}[t]
    \centering
    \includegraphics[width=0.7\linewidth]{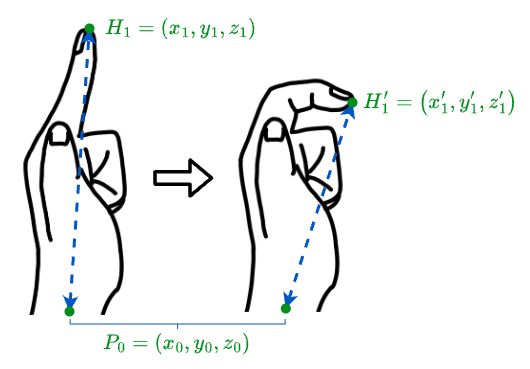}
    \caption{When users flex their index fingers, the system will dynamically update the coordinates of the tip of the index finger to calculate its distance to the wrist.}
    \Description{Two hand illustrations showing a user flexing their index finger. The system tracks the 3D coordinates of the fingertip and wrist before and after flexion, indicated by labeled points and dashed lines. The distance between the fingertip and the wrist is dynamically computed.}
    \label{fig:bezier_1}
\end{figure}

The gesture capture process begins by defining the initial positions of the wrist and fingertip (denoted as $\bm{P}_0$ and $\bm{H}_1$, respectively) when the finger is fully extended (see Figure~\ref{fig:bezier_1}). These positions are expressed in 3D Cartesian coordinates $(x, y, z)$, where $x$, $y$, and $z$ correspond to the horizontal, vertical, and depth dimensions. The maximum distance, $L_{\text{straight}}$, is calculated using the Euclidean distance formula:
\begin{equation}
    L_{\text{straight}} = \sqrt{(x_1-x_0)^2 + (y_1-y_0)^2 + (z_1-z_0)^2}
\end{equation}

As the finger flexes, this distance decreases to $L_{\text{bent}}$, and the curvature, $\kappa$, is computed as:
\begin{equation}
    \kappa = K_1 \cdot \left(\frac{L_{\text{straight}} - L_{\text{bent}}}{L_{\text{straight}}}\right)
\end{equation}
where $K_1 = 1.5$ is empirically determined to optimize interaction learnability and proved effective for the authors \rev{and during the in-lab pilot study with 4 participants}; however, this coefficient can be adjusted flexibly.

The initial point \( \bm{P}_{\text{0}} \) is positioned at the wrist, correlating with the virtual left hand's juncture. We establish \( \bm{v}_{\text{align}} \) as the unit vector extending parallel to the longitudinal axis of the forearm and \( \bm{v}_{\text{ortho}} \) as the unit vector perpendicular to the hand's plane. These vectors are instrumental in the determination of the locations for \( \bm{P}_{\text{1}} \) and \( \bm{P}_{\text{2}} \), which are modulated by \( \kappa \) and \( \ell \) as follows:

\begin{align}
    \bm{P}_{\text{1}} &= \bm{P}_{\text{0}} + \bm{v}_{\text{align}} \cdot \frac{1}{2} \cdot (1 + \kappa) \cdot \ell, \\
    \bm{P}_{\text{2}} &= \bm{P}_{\text{0}} + \ell \cdot \bm{v}_{\text{align}} + \kappa \cdot \ell \cdot \bm{v}_{\text{ortho}}.
\end{align}

Through the modulation of \( \kappa \) by the user's gesture of flexing their left index finger, the curvature of the Bézier curve is directly influenced. The dynamic modulation of the endpoint \( \bm{P}_{\text{2}} \) guarantees its alignment parallel to the wrist, thus facilitating an easy gestural interaction. Moreover, the calculated position of the control point \( \bm{P}_{\text{1}} \), derived from \( \kappa \), allows for the direct manipulation of the flexing trajectory of the curve, thereby synchronizing the Bézier Curve with the movements and curvature of the hand. Therefore, we have our quadratic Bézier Curve, which is defined by \( \bm{P}_{\text{0}} \), \( \bm{P}_{\text{1}} \), and \( \bm{P}_{\text{2}} \) (see Figure~\ref{fig:merged_system} (1)).

\subsection{Proximity-Matching Mechanism}

The forearm presents several compelling advantages for disambiguation tasks in VR. It is quick to access, benefits from strong proprioceptive awareness, and supports passive haptic feedback, allowing users to rely on body memory rather than constant visual confirmation~\cite{li2024onbodymenu,tsimbalistaia2025body}. Building on these findings, we developed a proximity-matching projection mechanism that dynamically identifies the four closest candidate objects—whether from a Linear Ray or Bézier Curve path—and projects them onto the user’s forearm. The decision to display four targets was informed by empirical findings from~\cite{li2024onbodymenu}, which demonstrated that limiting item size to the width of the forearm provided a comfortable and accurate selection experience. Expanding beyond this number risks reducing target size and increasing accidental activation. Thus, while the forearm's spatial constraints impose a natural limit, the projection of four carefully selected candidates balances efficiency, precision, and user comfort. This mechanism enables rapid and comfortable disambiguation by allowing users to select from spatially mapped options using proprioceptive cues and tactile feedback, regardless of whether the original selection path is linear or curved.

Our system computes the minimum Euclidean distance from a point $\bm{O}$ to a discretized Bézier Curve for efficient real-time interaction. The Bézier Curve is approximated using 20 linear segments to streamline computational processes. Each segment is defined by points on the curve:
\begin{equation}
    \bm{S}_i = \bm{B}\left(\frac{i}{n}\right), \quad i \in \{0, 1, 2, ..., n\}
\end{equation}
where $n = 20$, and $\bm{B}(t)$ denotes the curve parametrized by $t$.

The minimum distance, $d_{\text{min}}$, from the point $\bm{O}$ to the curve is computed as the smallest distance to any segment:
\begin{equation}
    d_{\text{min}} = \min_{i} \left( d(\bm{O}, [\bm{S}_i, \bm{S}_{i+1}]) \right)
\end{equation}
where $d(\bm{O}, [\bm{S}_i, \bm{S}_{i+1}])$ measures the Euclidean distance from $\bm{O}$ to the linear segment between $\bm{S}_i$ and $\bm{S}_{i+1}$.

\begin{figure*}[t]
    \centering
    \includegraphics[width=0.85\linewidth]{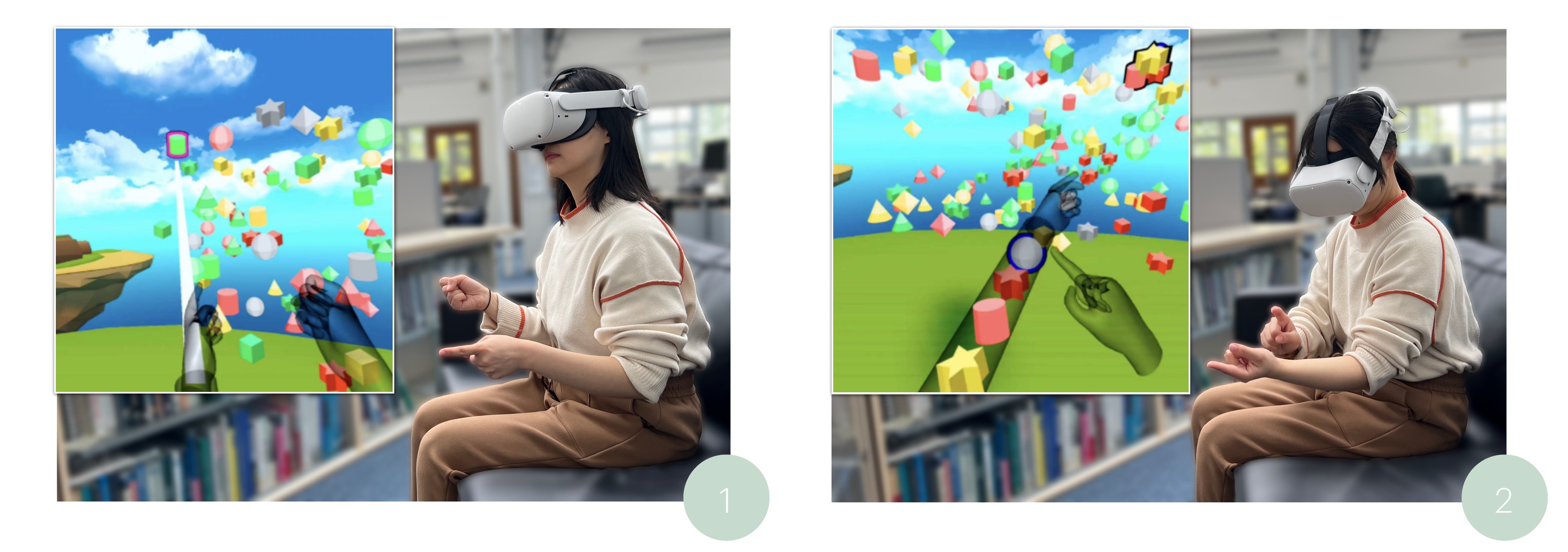}
    \caption{An example of a user using the mid-air surface for object selection (1) and the on-body surface for object selection (2).}
    \Description{Two side-by-side images showing a user wearing a VR headset performing object selection. In the first image, the user points into mid-air with a linear ray to select floating virtual objects. In the second image, the user interacts with a curved interface projected onto their forearm for on-body object selection. Each image includes an in-VR view showing the interaction from the user's perspective.}
    \label{fig:bezier_system}
\end{figure*}

The orthogonal projection $\bm{O}_{\text{proj}}$ of $\bm{O}$ onto each segment is critical for determining interaction points accurately and swiftly:
\begin{equation}
    \bm{O}_{\text{proj}} = \bm{S}_i + \lambda (\bm{S}_{i+1} - \bm{S}_i)
\end{equation}
where $\lambda$ is calculated as:
\begin{equation}
    \lambda = \left( \frac{(\bm{O} - \bm{S}_i) \cdot (\bm{S}_{i+1} - \bm{S}_i)}{\| \bm{S}_{i+1} - \bm{S}_i \|^2} \right)
\end{equation}

The computation of $\lambda$ is further optimized using Unity's built-in functionality:
\begin{equation}
    \lambda(\bm{S}_{i+1} - \bm{S}_i) = \textit{Vector3.Project}(\bm{O} - \bm{S}_i, \bm{S}_{i+1} - \bm{S}_i)
\end{equation}

This methodology leverages Unity's efficient vector operations\footnote{https://docs.unity3d.com/6000.0/Documentation/ScriptReference/\\Vector3.Project.html} to ensure that the real-time computation of $\lambda$ is both accurate and responsive, which is essential for maintaining the fluidity and precision of user interactions in our system (see Figure~\ref{fig:merged_system} (3)).

Finally, user selection activation and locking mechanisms are controlled through hand gestures as well. Activation occurs when the right middle finger is bent, which initiates the appearance of the ray and enables the dynamic mapping of objects onto the left forearm in real time. Conversely, straightening the right middle finger locks the selection, causing the ray to disappear and stabilizing the mapping of objects on the left forearm. This facilitates a more straightforward selection of the target object by the user.

\section{User Study}
To evaluate our proposed on-body disambiguation mechanism across different selection paradigms, we conducted a user study with 24 participants. Our aim was to examine how these design choices influence task performance and user experience. We framed our study around the following research questions:

\begin{itemize}
  \item \textbf{RQ1:} How does the proposed on-body disambiguation mechanism affect selection performance (task time and error rate) across paradigms?
  \item \textbf{RQ2:} What is the impact of the on-body mechanism on subjective workload, usability, and experience?
  \item \textbf{RQ3:} How does the Bézier interaction paradigm compare to Linear Ray-casting in terms of performance and user perception, with and without the on-body mechanism?
\end{itemize}

Based on prior work in disambiguation and curved interaction, we formulated the following hypotheses:

\begin{itemize}
  \item \textbf{H1:} The on-body disambiguation mechanism will significantly reduce selection errors and improve perceived performance.
  \item \textbf{H2:} On-body interaction will reduce physical and perceived workload and enhance subjective preference and hedonic quality.
  \item \textbf{H3:} The Bézier paradigm will improve performance in cluttered scenes but may increase task time or effort compared to Linear Ray-casting.
\end{itemize}

\subsection{Conditions and Technique Design}

\rev{We have four conditions: \textsc{Interaction Paradigm} (Bézier Curve vs. Linear Ray) $\times$ \textsc{Interaction Medium} (On-body vs. Mid-air). This resulted in four experimental conditions, enabling us to investigate how the proposed on-body disambiguation mechanism and the curve-based selection paradigm perform independently and in combination.} The on-body disambiguation mechanism, based on proximity-matching projection, was enabled only in the on-body conditions. To ensure stability and reduce hand tremor effects, we adopted a bimanual interaction pattern: participants used their non-dominant (e.g., left) hand to indicate a selection gesture, while confirming selection with their dominant (e.g., right) hand. The confirmation was triggered by extending the index finger from a resting fist posture, which was tracked in real time by the VR headset.

\paragraph{Mid-air Linear Ray}  
This condition represents the default selection behavior in most commercial VR systems. Participants pointed at targets in mid-air using a straight ray projected from their left index finger, and confirmed the selection with a gesture from their right hand.

\paragraph{Mid-air Bézier Curve}  
This condition replaces the Linear Ray with a dynamic Bézier Curve generated based on the bending angle of the participant’s index finger. The control structure of the Bézier path enables indirect or occluded target access while maintaining mid-air interaction. Selection was confirmed using the same bimanual gesture as the Linear Ray condition.

\paragraph{On-body Linear Ray}  
In this condition, we overlaid the four nearest ray intersections onto the user’s forearm using our proximity-matching projection mechanism. Participants viewed the candidates on their forearm, identified the projected target, and touched it with their right hand to confirm selection. The visual ray itself remained linear, but disambiguation occurred via on-body interaction.

\paragraph{On-body Bézier Curve}  
Similar to the previous condition, this variant used the Bézier Curve as the underlying selection path, but with the same on-body disambiguation mechanism projecting the four closest intersections onto the forearm. Participants confirmed selection by touching the target as projected on their bodies.

\begin{figure}[t]
    \centering
    \includegraphics[width=0.75\linewidth]{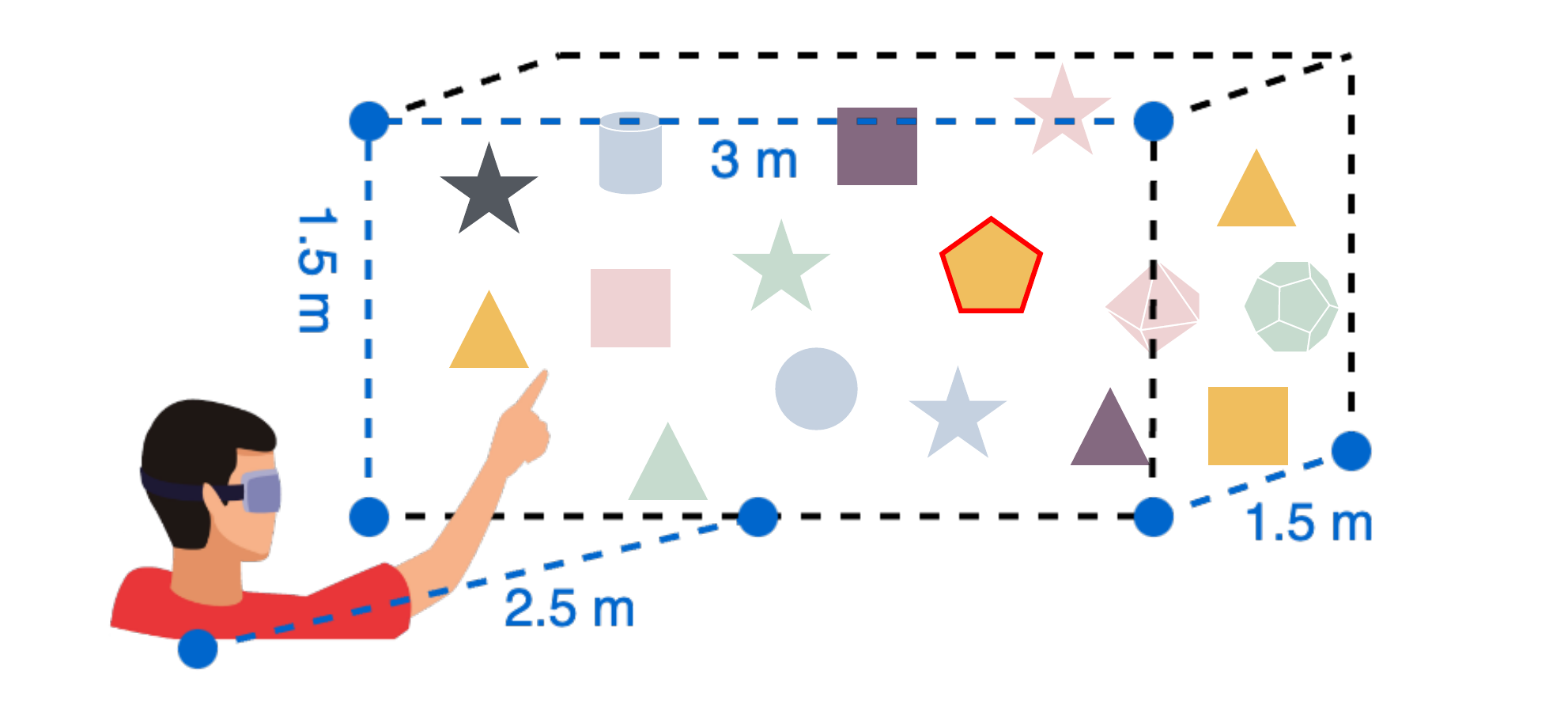}
    \caption{In each round of the user study, 64 icons of various colors and shapes are randomly distributed within a designated space measuring 1.5 $\times$ 1.5 $\times$ 3 meters. Participants stand 2.5 meters away from the front edge of this space. A single icon is highlighted as the target for selection.}
    \Description{Illustration of a user standing 2.5 meters in front of a 3D interaction space measuring 1.5 by 1.5 by 3 meters. The space contains 64 icons of different colors and shapes scattered randomly. One icon is highlighted to indicate it as the target for selection.}
    \label{fig:setup}
\end{figure}

\subsection{Study Design}

While the mid-air condition supports single-target ray intersection, the on-body condition displays up to four candidates, reflecting a practical implementation bounded by forearm size and proprioceptive clarity, rather than a strict symmetry in design.

Each trial began with 64 virtual objects randomly distributed within an invisible 3D volume measuring 1.5~m (height) × 1.5~m (depth) × 3~m (width), positioned approximately 2.5~m in front of the participant. One object was randomly highlighted as the target. The objects were distinct icons varying in color and shape to facilitate visual recognition while avoiding icon familiarity biases. A schematic of this layout is shown in Figure~\ref{fig:setup}.

Although we did not manipulate spatial density or occlusion as formal independent variables, our use of 64 virtual objects within a confined 3D volume inherently induced high visual clutter. This setup frequently led to target overlap and partial occlusion, requiring users to resolve ambiguity and exercise fine control during selection. The design was motivated by common MR scenarios (e.g., virtual tool palettes, dense object fields, or data-rich workspaces) where disambiguation and trajectory flexibility are crucial. However, we acknowledge that formal manipulation of occlusion and density remains an important direction for future work to validate the generalizability of our findings across diverse spatial conditions.

Each participant completed 30 trials per condition, resulting in a total of 2,880 trials across the study (2 paradigms × 2 media × 30 trials × 24 participants). This design allowed us to evaluate not only the main effects of the interaction paradigm and interaction medium but also any interaction effects between them. This user study was approved by our Department Ethics Application.

\subsection{Participants and Apparatus}

24 participants (9 female, 15 male) were recruited from a local university. Participants ranged in age from 18 to 29 years ($M = 22.46$, $SD = 1.60$). All participants were right-handed and reported normal or corrected-to-normal vision. Prior VR experience was assessed using a 7-point Likert scale, where 1 indicated no prior experience and 7 indicated expert-level familiarity ($M = 4.14$, $SD = 2.35$). All participants used a Meta Quest 2 headset with hand-tracking enabled. A small monetary compensation was provided upon completion of the study. 

\subsection{Measures}

To evaluate the effects of interaction paradigm and interaction medium, we collected both objective and subjective measures. Task performance was assessed using task completion time (in seconds) and the number of selection errors (i.e., incorrect target selections). Subjective experience was captured through several validated instruments: perceived workload was measured using the raw NASA-TLX~\cite{hart_nasa-task_2006}, usability via the System Usability Scale (SUS)~\cite{brooke1996sus}, and user experience through the UEQ-S~\cite{schrepp2017ueq}. We also measured enjoyment using a single 7-point Likert item. At the end of the session, participants were asked to rank their final preferences across the four conditions using a 5-point Likert scale.

\subsection{Procedure}

\begin{table*}[ht]
\centering
\caption{
Descriptive statistics (Mean and SD) for all dependent measures across main effects: Medium (On-Body and Mid-Air) and Paradigm (Bézier Curve and Linear Ray).
}
\label{tab:main_effects}
\begin{tabular}{lcccc}
\toprule
\textbf{Measure} & \textbf{On-Body} & \textbf{Mid-Air} & \textbf{Bézier Curve} & \textbf{Linear Ray} \\
\midrule
\textit{Performance Metrics} \\
Task Completion Time & 198.14 (51.04) & 180.22 (44.52) & 199.95 (54.45) & 178.41 (39.37) \\
Avg Time per Attempt & 5.04 (1.43) & 3.93 (0.96) & 4.69 (1.52) & 4.28 (1.09) \\
Error Attempts       & 10.23 (7.48) & 17.23 (11.58) & 14.19 (10.41) & 13.27 (10.31) \\
\midrule
\textit{NASA-TLX} \\
Mental Demand     & 2.77 (1.48) & 3.12 (1.71) & 3.10 (1.64) & 2.79 (1.56) \\
Physical Demand   & 3.33 (1.68) & 4.15 (1.77) & 3.77 (1.70) & 3.71 (1.84) \\
Temporal Demand   & 3.08 (1.43) & 3.19 (1.63) & 3.17 (1.40) & 3.10 (1.65) \\
Performance       & 5.31 (1.40) & 4.71 (1.30) & 4.92 (1.35) & 5.10 (1.42) \\
Effort            & 3.52 (1.62) & 3.92 (1.57) & 3.67 (1.56) & 3.77 (1.65) \\
Frustration       & 2.92 (1.57) & 3.29 (1.65) & 3.21 (1.69) & 3.00 (1.54) \\
\midrule
\textit{UEQ-S} \\
Pragmatic Quality & 1.13 (1.09) & 0.91 (1.32) & 1.06 (1.10) & 0.98 (1.32) \\
Hedonic Quality   & 1.15 (1.00) & 0.31 (1.46) & 0.96 (1.13) & 0.50 (1.45) \\
Overall Score     & 1.14 (0.94) & 0.61 (1.21) & 1.01 (1.00) & 0.74 (1.21) \\
\midrule
\textit{SUS} & 70.68 (16.73) & 73.02 (16.19) & 70.26 (18.35) & 73.44 (14.25) \\
\midrule
\textit{Final Preference} & 3.38 (1.20) & 3.52 (1.24) & 3.44 (1.20) & 3.46 (1.24) \\
\textit{Enjoyment} & 5.04 (1.47) & 5.08 (1.47) & 4.98 (1.52) & 5.15 (1.41) \\
\bottomrule
\end{tabular}
\end{table*}

\begin{table*}[ht]
\centering
\caption{
Descriptive statistics (Mean and SD) for all dependent measures across combined Medium × Paradigm conditions.
}
\label{tab:combined_conditions}
\begin{tabular}{lcccc}
\toprule
\textbf{Measure} & \textbf{On-body Bézier Curve} & \textbf{On-body Linear Ray} & \textbf{Mid-air Bézier Curve} & \textbf{Mid-air Linear Ray} \\
\midrule
\textit{Performance Metrics} \\
Task Completion Time & 211.40 (62.17) & 184.89 (33.04) & 188.51 (43.80) & 171.93 (44.59) \\
Avg Time per Attempt & 5.45 (1.68) & 4.63 (1.00) & 3.93 (0.85) & 3.92 (1.08) \\
Error Attempts       & 9.46 (6.02) & 11.00 (8.76) & 18.92 (11.77) & 15.54 (11.38) \\
\midrule
\textit{NASA-TLX} \\
Mental Demand     & 2.92 (1.50) & 2.62 (1.47) & 3.29 (1.78) & 2.96 (1.65) \\
Physical Demand   & 3.42 (1.67) & 3.25 (1.73) & 4.12 (1.70) & 4.17 (1.88) \\
Temporal Demand   & 3.12 (1.36) & 3.04 (1.52) & 3.21 (1.47) & 3.17 (1.81) \\
Performance       & 5.17 (1.40) & 5.46 (1.41) & 4.67 (1.27) & 4.75 (1.36) \\
Effort            & 3.71 (1.73) & 3.33 (1.52) & 3.62 (1.41) & 4.21 (1.69) \\
Frustration       & 3.12 (1.68) & 2.71 (1.46) & 3.29 (1.73) & 3.29 (1.60) \\
\midrule
\textit{UEQ-S} \\
Pragmatic Quality & 1.11 (0.91) & 1.15 (1.26) & 1.01 (1.28) & 0.81 (1.38) \\
Hedonic Quality   & 1.15 (1.06) & 1.16 (0.96) & 0.77 (1.19) & -0.16 (1.58) \\
Overall Score     & 1.13 (0.87) & 1.15 (1.03) & 0.89 (1.12) & 0.33 (1.25) \\
\midrule
\textit{SUS} & 70.31 (18.96) & 71.04 (14.58) & 70.21 (18.13) & 75.83 (13.81) \\
\midrule
\textit{Final Preference} & 3.46 (1.04) & 3.29 (1.36) & 3.42 (1.38) & 3.63 (1.08) \\
\textit{Enjoyment} & 5.04 (1.46) & 5.04 (1.52) & 4.92 (1.61) & 5.25 (1.33) \\
\bottomrule
\end{tabular}
\end{table*}

Participants began with a brief orientation, during which they were introduced to the goals and the interaction techniques used. After providing informed consent, they completed a demographic questionnaire. Before each condition, participants were given a five-minute training session to familiarize themselves with the respective interaction technique. Each experimental condition involved a standardized pointing task designed to evaluate selection performance and user experience. The order of conditions was counterbalanced using a Latin square design to mitigate potential order effects. After completing each condition, participants filled out a post-condition questionnaire capturing subjective ratings on workload, usability, enjoyment, and user experience. Upon finishing all conditions, they completed a final preference questionnaire and participated in a semi-structured interview to provide qualitative feedback. Each session lasted approximately one hour per participant.

\section{Results}

We conducted a two-way repeated-measures analysis to examine the effects of \textsc{Medium} (On-body vs. Mid-air) and \textsc{Paradigm} (Linear Ray vs. Bézier Curve) on each dependent variable. Normality was assessed using the Shapiro--Wilk test within each condition. When all conditions met normality assumptions, we applied a repeated-measures ANOVA; otherwise, we used the Aligned Rank Transform (ART)~\cite{wobbrock2011art} to enable non-parametric factorial analysis. If a significant interaction was found, we performed simple effects analyses by comparing factor levels within each condition: we used paired $t$-tests or Wilcoxon signed-rank tests depending on the normality of the difference scores, applying Holm correction to all $p$-values. For ART-based models, we used ART-Contrasts (ART-C)~\cite{elkin2021artc} to perform follow-up comparisons. When no interaction was found, we directly interpreted main effects; no post-hoc tests were required as each factor had only two levels.

\subsection{Task Performance}

\paragraph{Task Completion Time.}
A significant main effect of \textsc{Paradigm} was found for \textit{Task Completion Time} ($F(1, 23) = 8.84$, $p = .007$), with faster performance in the Linear condition ($M = 178.41$, $SD = 39.37$) than in the Bézier condition ($M = 199.95$, $SD = 54.45$). No significant effects were observed for \textsc{Medium} ($F(1, 23) = 3.38$, $p = .079$) or the interaction between factors ($F(1, 23) = 0.60$, $p = .448$).

\paragraph{Average Time per Attempt.}
A significant interaction was found for \textit{Average Time per Attempt} ($F(1, 23) = 5.74$, $p = .025$). Simple effects analysis using paired $t$-tests with Holm correction showed that Bézier took significantly longer than Linear in the On-body condition ($t(23) = 2.79$, $p = .031$), and On-body took significantly longer than Mid-air in both the Bézier ($t(23) = -4.13$, $p = .0016$) and Linear ($t(23) = -2.39$, $p = .050$) conditions. No other comparisons were significant.

\paragraph{Error Attempts.}
A significant main effect of \textsc{Medium} was found for \textit{Error Attempts} using ART analysis ($F(1, 69) = 13.59$, $p < .001$), with fewer errors in the On-body condition ($M = 10.23$, $SD = 7.48$) than in the Mid-air condition ($M = 17.23$, $SD = 11.58$). No significant effects were observed for \textsc{Paradigm} ($F(1, 69) = 0.85$, $p = .36$) or the interaction between factors ($F(1, 69) = 1.23$, $p = .27$). Follow-up ART-Contrasts confirmed the main effect of \textsc{Medium} ($p = .0004$).

\subsection{Perceived Workload (NASA-TLX)}

\paragraph{Physical Demand.}
A significant main effect of \textsc{Medium} was found for \textit{Physical Demand} ($F(1, 69) = 9.87$, $p = .002$), with lower ratings reported in the On-body condition ($M = 3.33$, $SD = 1.68$) compared to Mid-air ($M = 4.15$, $SD = 1.77$). No significant effects were found for \textsc{Paradigm} ($F(1, 69) = 0.11$, $p = .737$) or the interaction term ($F(1, 69) = 0.05$, $p = .819$). Follow-up ART-Contrasts confirmed the difference between interaction media ($p = .0025$).

\paragraph{Performance.}
A significant main effect of \textsc{Medium} was also found for \textit{Performance} ($F(1, 69) = 7.72$, $p = .007$), with higher ratings in On-body conditions ($M = 5.31$, $SD = 1.40$) than in Mid-air ($M = 4.71$, $SD = 1.30$). No significant effects were found for \textsc{Paradigm} ($F(1, 69) = 0.72$, $p = .398$) or the interaction ($F(1, 69) = 0.10$, $p = .758$). ART-Contrasts confirmed the effect of \textsc{Medium} ($p = .007$).

\paragraph{Other Workload Dimensions.}
No significant effects were found for \textit{Mental Demand} ($F(1, 69) = 1.23$, $p = .270$), \textit{Temporal Demand} ($F(1, 69) = 0.03$, $p = .854$), \textit{Effort} ($F(1, 69) = 1.41$, $p = .240$), or \textit{Frustration} ($F(1, 69) = 1.17$, $p = .283$). These results suggest that while the disambiguation mechanism significantly reduced physical demand and improved performance ratings, it did not significantly affect other aspects of workload.

\subsection{Usability and User Experience}

\paragraph{System Usability.}
No significant effects were found for \textsc{Medium} ($F(1, 23) = 1.06$, $p = .313$), \textsc{Paradigm} ($F(1, 23) = 2.01$, $p = .170$), or their interaction ($F(1, 23) = 1.25$, $p = .275$). Simple effects comparisons were all non-significant after Holm correction, indicating no reliable difference in perceived usability across conditions. SUS scores across all conditions were above 70 on average, suggesting an overall positive usability impression.

\paragraph{UEQ: Pragmatic Quality.}
No significant effects were found for \textsc{Medium} ($F(1, 23) = 0.65$, $p = .428$), \textsc{Paradigm} ($F(1, 23) = 0.25$, $p = .624$), or their interaction ($F(1, 23) = 0.41$, $p = .529$). Simple effects comparisons were all non-significant.

\paragraph{UEQ: Hedonic Quality.}
We observed a significant interaction for \textit{Hedonic Quality} ($F(1, 23) = 7.13$, $p = .014$), as well as significant main effects of \textsc{Medium} ($F(1, 23) = 10.14$, $p = .004$) and \textsc{Paradigm} ($F(1, 23) = 5.66$, $p = .026$). Simple effects analysis revealed that Linear Ray in the Mid-air condition was rated significantly lower than Bézier ($t(23) = 2.92$, $p = .023$), while no other comparisons reached significance after Holm correction.

\paragraph{UEQ: Overall Score.}
A main effect of \textsc{Medium} was found for \textit{Overall Score} ($F(1, 23) = 4.82$, $p = .039$), with higher ratings for On-body interaction ($M = 1.14$, $SD = 0.94$) compared to Mid-air ($M = 0.61$, $SD = 1.21$). A main effect of \textsc{Paradigm} was also observed ($F(1, 23) = 4.70$, $p = .041$), with higher scores for Bézier ($M = 1.01$, $SD = 1.00$) than Linear Ray ($M = 0.74$, $SD = 1.21$). The interaction was marginally significant ($F(1, 23) = 3.29$, $p = .083$), and a simple effects comparison indicated Bézier was preferred over Linear in the Mid-air condition ($t(23) = 2.80$, $p = .041$).

\subsection{Final Preference and Enjoyment}

\paragraph{Final Preference.}
A significant interaction was found for \textit{Final Preference} ($F(1, 23) = 12.31$, $p = .0019$). Simple effects analysis using Wilcoxon signed-rank tests with Holm correction revealed that participants preferred Bézier over Linear in the On-body condition ($W = 41.5$, $p = .0338$), and preferred On-body over Mid-air in the Bézier condition ($W = 45.5$, $p = .0382$). Other comparisons were not significant.

\paragraph{Enjoyment.}
No significant effects were found for \textsc{Medium} ($F(1, 69) = 0.05$, $p = .828$), \textsc{Paradigm} ($F(1, 69) = 0.12$, $p = .727$), or their interaction ($F(1, 69) = 0.17$, $p = .678$). \textit{Enjoyment} scores were consistently high across all conditions (all $M > 4.9$), indicating a generally positive user experience regardless of interaction technique.

\subsection{Semi-Structured Interview}

To complement our quantitative findings, we conducted a semi-structured interview with each participant to gather qualitative insights into their experiences across different \textit{interaction paradigms} (Linear Ray vs. Bézier Curve) and \textit{interaction media} (Mid-air vs. On-body). We analyzed the case study data using an inductive thematic analysis~\cite{braun2006using}, which revealed three major themes: \textit{Usability of Interaction Paradigms}, \textit{Experiences with Interaction Media}, as well as \textit{Practical Challenges} encountered during selection. 

\paragraph{Usability of Interaction Paradigms.}  
Participants expressed their preferences between the Bézier Curve and Linear Ray paradigms. The Bézier Curve was often described as more powerful in dense scenes due to its spatial flexibility. Several users noted it enabled precise control when selecting occluded targets or navigating cluttered environments. However, this control came at a cost: many participants reported increased fatigue from the finger bending required to shape the curve. As P6 observed, \textit{``the Bézier Curve required too much finger flexion, which became tiring during extended use.''} P2 also remarked, \textit{``the curved one is not as good as the straight one; it cannot bend to the degree I want and feels less responsive.''} In contrast, the Linear Ray paradigm was appreciated for its simplicity and alignment with natural pointing gestures. P7 summarized this trade-off well: \textit{``the ray worked well for simple tasks but struggled with occluded targets or dense objects.''}

\paragraph{Experiences with Interaction Media.}  
Participants generally preferred the on-body interaction medium for its tactile qualities and proprioceptive support. The physicality of touching the forearm provided a strong sense of control and confirmation. As P10 shared, \textit{``the physical contact gave a strong sense of control and confirmation.''} On-body interaction was described as more immersive and embodied, helping users to anchor their actions in space. However, this technique also introduced challenges. For instance, P3 noted that \textit{``the need for high recognition accuracy with both hands''} occasionally led to input errors during confirmation. Mid-air interaction, by contrast, was praised for its speed and minimal occlusion of the virtual scene, but was seen as lacking physical feedback. Some participants described mid-air techniques as more ``floaty'' or ambiguous, especially when selecting between closely spaced targets.

\paragraph{Challenges with the Proximity-Matching Mechanism.}  
Participants also commented on the limitations of the on-body disambiguation mechanism, particularly in relation to occlusion handling and camera positioning. Several users experienced difficulty when the projected candidates did not align well with the intended target. As P2 stated, \textit{``sometimes it cannot be selected when occluded,''} while P5 mentioned, \textit{``the selection often depends on the user’s position.''} P8 noted that \textit{``the ray is not very responsive and is sometimes blocked,''} and P4 highlighted a technical limitation: \textit{``when my finger and the camera are parallel, the mechanism sometimes fails to register properly.''} These comments point to important areas for improvement, including enhanced occlusion resolution, robustness to hand orientation, and greater responsiveness under varied viewpoints.

\section{Discussion}

This study examined how our proposed on-body disambiguation mechanism and curve-based interaction paradigm affect selection performance, subjective workload, usability, and user preference in immersive environments. We addressed three research questions (\textbf{RQ1}–\textbf{RQ3}) and tested corresponding hypotheses (\textbf{H1}–\textbf{H3}).

\paragraph{RQ1: Effects on Task Performance.}
Our findings strongly support \textbf{H1}. On-body interaction significantly reduced selection errors across both interaction paradigms, validating the efficacy of our disambiguation mechanism in visually dense environments. However, this improvement came at a cost of increased task time, particularly when combined with the Bézier paradigm. These results partially support \textbf{H3}, which predicted that Bézier Curves would offer higher precision at the expense of efficiency. While Bézier enabled more flexible target access, especially in occluded settings, it introduced a speed-accuracy trade-off, exacerbated by the additional steps in on-body selection.

\paragraph{RQ2: Impact on Workload, Usability, and Experience.}
We found partial support for \textbf{H2}. On-body interaction improved perceived performance and reduced physical demand, especially compared to Mid-air input. It also yielded higher hedonic and overall UEQ scores, suggesting enhanced experiential quality. However, we did not observe significant differences in overall usability (SUS), enjoyment, or effort, nor consistent effects of paradigm alone. These results indicate that the benefits of on-body interaction lie more in its perceived expressiveness and control rather than in reducing workload across all dimensions. Notably, the Bézier paradigm did not independently impact most subjective measures, suggesting its value may depend on pairing with supportive mechanisms like our on-body approach.

\paragraph{RQ3: Subjective Preference and Trade-offs.}
User preferences further reinforced \textbf{H2} and \textbf{H3}. Participants favored On-body techniques, citing tactile grounding and spatial alignment as key to confident selection. In contrast, responses to the Bézier paradigm were mixed: while some users valued its flexibility for navigating occlusion, others found it less intuitive or physically taxing. These responses suggest that Bézier interaction is best applied in context-sensitive scenarios requiring spatial adaptability rather than as a universal alternative to Linear Ray casting. Taken together, our results highlight that precision-enhancing mechanisms are most effective when thoughtfully integrated with interaction paradigms.

\paragraph{Design Implications.} \rev{Our results suggest that Bézier and Linear should be seen as complementary rather than competing techniques. Linear rays remain the fastest and least fatiguing option for simple, low-density scenes, while Bézier is valuable in cluttered or occluded layouts where direct pointing fails. A practical strategy is to treat Linear as the default and invoke Bézier adaptively when ambiguity or occlusion is detected.}

Our findings offer actionable insights for immersive interaction design. First, the proposed on-body disambiguation mechanism significantly improves selection accuracy and is well-received by users across interaction paradigms. Designers should consider invoking such mechanisms dynamically, particularly in scenes with dense or overlapping objects, to support confident and precise target selection.

Second, Bézier interaction serves as a flexible alternative to linear raycasting, especially in occluded or spatially complex environments. Our Bézier curve generation algorithm is lightweight and replicable for controller-based systems. For example, in controller mode, the trigger button depth can control the ray curvature $\kappa$:
\begin{equation}
    \kappa = K_2 \cdot d_{trigger}
\end{equation}

where $K_2$ is empirically set to approximately 0.2, and $d_{trigger}$ denotes the trigger actuation depth from 0 (no press) to 1 (full press). This mapping aligns the curvature range with that observed in hand-tracking mode, enabling consistent behavior across input modalities. That said, designers should be aware of the physical and cognitive demands introduced by explicit curve shaping. A promising direction is to infer curvature implicitly from finger motion or to blend Bézier and linear rays into a unified, context-aware model.

Third, the limitations of our proximity-based matching, such as reduced accuracy under occlusion or positional misalignment, highlight opportunities for improvement. Enhancing sensing robustness and projection responsiveness will be key to supporting reliable on-body interaction across diverse scenarios.

\section{Limitations and Future Work}

While our proposed on-body disambiguation mechanism yielded promising results, several limitations merit discussion and suggest fruitful directions for future work. First, our proximity-matching system approximates the Bézier curve using discretized linear segments to ensure real-time responsiveness. While this approach proved effective in practice, it may introduce minor approximation errors, especially when targets lie close to but not directly on a sampled segment. Future work could explore adaptive sampling resolution or incorporate numerically optimized distance metrics directly on continuous Bézier curves.

Second, although we evaluated our techniques in a visually dense and partially occluded environment, we did not manipulate visual complexity or occlusion level as formal independent variables. The use of a fixed high-density layout enabled controlled comparisons across conditions but limits generalizability to broader spatial configurations. Future studies could vary spatial clutter, occlusion severity, and object distribution to systematically assess scalability and robustness.

Third, our study design introduces a structural asymmetry between mid-air and on-body conditions: the on-body disambiguation mechanism presents up to four candidate targets via forearm projection, while the mid-air technique supports direct selection of a single target. While this reflects practical design trade-offs in current MR systems, it also introduces potential confounds between interaction medium and disambiguation strategy. We acknowledge that this asymmetry makes it difficult to attribute performance differences solely to the medium or technique. Future work should explore more symmetric baselines—for example, mid-air radial menus or spatial previews—to isolate the effects of projection modality and selection complexity.

Finally, the current on-body design supports disambiguation for only up to four candidates, constrained by forearm width and proprioceptive clarity. While this number was grounded in prior findings, it may not scale to more complex environments or larger candidate sets. Future extensions could explore multi-region on-body interaction (e.g., bilateral arms), as well as the integration of multimodal cues (e.g., audio or gaze-guided highlighting) to support larger disambiguation sets without sacrificing clarity or speed.

\section{Conclusion}

We presented two complementary techniques to enhance object selection in immersive environments: a Bézier Curve-based selection paradigm guided by finger curvature, and an on-body disambiguation mechanism that projects candidate objects onto the user’s forearm. These techniques were evaluated in a controlled $2 \times 2$ factorial study crossing interaction paradigm (Bézier Curve vs. Linear Ray) and interaction medium (Mid-air vs. On-body).

Our results show that on-body disambiguation significantly improves selection accuracy, reduces physical workload, and increases user preference in dense or occluded scenes. The forearm interface supports tactile grounding and proprioceptive clarity, enabling confident disambiguation. The Bézier paradigm offers flexible access to occluded targets, though with trade-offs in task time and demand. Together, our findings demonstrate how on-body disambiguation and curved interaction paradigms can be jointly leveraged to improve precision and user experience in 3D interfaces. These insights inform the design of next-generation immersive systems that are more expressive, adaptive, and human-centered.

\begin{acks}
Xiang Li is supported by the China Scholarship Council (CSC) International Cambridge Scholarship (No. 202208320092). Per Ola Kristensson is supported by the EPSRC (EP/W02456X/1). We would like to thank Yuanxin Guo (University of Toronto), Yuzheng Chen (Lancaster University), and Wei He (HKUST(GZ)) for their feedback.
\end{acks}

\bibliographystyle{ACM-Reference-Format}
\bibliography{bezier}

\appendix
\vfill
\section{Pseudocode for Generating a Bézier Curve \& Find the Closest Objects}
\begin{algorithm}[b]
\caption{Generate a Bézier Curve}
\label{algo:maaip_algorithm}
\begin{algorithmic}[1]
\REQUIRE Constant coefficient $K$; terminal points $\bm{P}_0$
\ENSURE Quadratic Bézier Curve
\STATE $L_{\text{bent}} \leftarrow$ distance($\bm{H}_1^{'}$, $\bm{P}_0$)
\WHILE{$L_{\text{bent}} < L_{\text{straight}}$}
    \STATE $L_{\text{bent}} \leftarrow$ distance($\bm{H}_1^{'}$, $\bm{P}_0$)
    \STATE $\kappa \gets K \cdot \frac{{L_{\text{straight}} - L_{\text{bent}}}}{{L_{\text{straight}}}}$
    \STATE $\bm{P}_1 \leftarrow \bm{P}_0 + v_{\text{align}} \cdot \frac{1}{2} \cdot (1 + \kappa) \cdot \ell $
    \STATE $\bm{P}_2 \leftarrow \bm{P}_0 + \ell  \cdot v_{\text{align}} + \kappa \cdot \ell  \cdot v_{\text{ortho}}$
     \FOR{$t \leq 1$}
        \STATE $c_0 \gets (1 - t)^2$
        \STATE $c_1 \gets 2 \cdot (1 - t) \cdot t$
        \STATE $c_2 \gets t^2$
        \STATE $\bm{B}(t) \gets c_0 \cdot \bm{P}_0 + c_1 \cdot \bm{P}_1 + c_2 \cdot \bm{P}_2$
    \ENDFOR
\ENDWHILE
\end{algorithmic}
\end{algorithm}

\begin{algorithm}[b]
\caption{Find Closest Objects}
\label{algo:closest_projections}
\begin{algorithmic}[1]
\REQUIRE Total $n$ segments, Total $m$ objects
\ENSURE Closest distance $d$ for each object; Return the closest 4 objects
\WHILE{Mapping is not done}
\FOR{$j = 1$ to $m$}
    \STATE $d_{\text{min},j} \gets \infty$
    \FOR{$i = 1$ to $n$}
        \STATE $\bm{O}_{\text{proj}} \gets \bm{S}_i + Vector3.Project(\bm{O}_j - \bm{S}_i, \bm{S}_{i+1} - \bm{S}_i)$
        \STATE $d \gets \text{distance}(\bm{O}_{\text{proj}}, \bm{O}_j)$
      
        \IF{$d < d_{\text{min},j}$}
            \STATE $d_{\text{min},j} \gets d$
        \ENDIF
    \ENDFOR
\ENDFOR
\STATE Sort objects by $d_{\text{min},j}$ in ascending order
\STATE Return the first 4 objects
\ENDWHILE
\end{algorithmic}
\end{algorithm}

\end{document}